\documentclass[aip,reprint]{revtex4-1}

\usepackage{datetime}
\usepackage{graphicx}

\begin{document}

\title{Molecular motions in glass and supercooled states of clotrimazole studied by broadband dielectric spectroscopy}

\author{N. S. K. Kumar}
\affiliation{Department of Physics, School of Physical, Chemical and Applied Sciences, Pondicherry University, Puducherry-605014, India}

\author{G. Govindaraj}
\email[Corresponding author, Email:]{ ggraj\_7@yahoo.com}
\affiliation{Department of Physics, School of Physical, Chemical and Applied Sciences, Pondicherry University, Puducherry-605014, India}

\author{U. Sailaja}
\affiliation{Department of Physics, M.E.S. Keveeyam College, Valancherry, Malappuram, Kerala-676552,  India}

\date{\today}

\begin{abstract}
The molecular mobility of glassy and supercooled liquid states of clotrimazole is studied using broadband dielectric spectroscopy for a wide range of temperatures and frequency. The dielectric loss data of clotrimazole below T$_{g}$, do not show well resolved relaxation. Above T$_{g}$, dielectric loss show the structural $\alpha$-relaxation which is reformed to shoulder like shape at high frequencies due to the secondary relaxation. The relaxation time of $\alpha$-process, $\tau_{\alpha}$ follows  Vogel-Fulcher-Tammann equation. The glass transition temperature, $T_{g}$=296K and fragility index, m=50 are obtained from the thermal behaviour of the $\alpha$-process. At temperature near T$_{g}$ the dielectric loss has Kohlrausch-Williams-Watts stretch exponential parameter $\beta_{KWW}$=0.56. The primitive relaxation time obtained from the coupling model coincides with the experimentally observed secondary relaxation of clotrimazole. Hence the secondary relaxation is considered to be the Johari-Goldstein (JG) process which may be the precursor $\alpha$-process. The value of exponent $\beta_{KWW}$ increases with increase of temperature and scaled dielectric loss do not form as single master curve. The structural relaxation $\alpha$-relaxation and JG $\beta$-relaxation process and its effects in clorimazole is discussed.
\end{abstract}


\maketitle 

\section{Introduction}
Many of the active pharmaceutical ingredients are poorly water soluble and having low bioavailability. Improving the solubility and bioavailability and hence reduce the risk of usage of large dosage to produce desirable effect in patients, is one of the prime aspect of pharmaceutical research\cite{AmorpPharm-2002}. Producing the drug in amorphous form is one of the method to improve the enhance dissolution rate and improve bioavailability. But amorphous materials are metastable having greater tendency for devitrification even it is stored below glass transition temperature\cite{AlieMolMobPharm-2004,StabilityPharm-1998}. Key factor of devitrification and reduced shelf life is the higher molecular mobility of amorphous pharmaceutical\cite{MolMobPharm-2010,MolMobPharm-2012,MolMobPharm-2002}.
\par
Broadband dielectric spectroscopy is one of the suitable tool to investigate dielectric properties of various kinds of materials- liquids, glasses, polymers and even disordered semiconductors\cite{BDS2002,KKPhysicaB}. The dielectric relaxation of the glass and supercooled liquid states of pharmaceutical drug gives information about the molecular mobility in this state\cite{DrugDelRev2015,SailajaKetoprofen,SailajaFenofibrate}. The knowledge of the molecular mobility of the amorphous pharmaceutical may useful to avoid devitrification and increase the shelf life of the material. One of the explanation on possible origin of devitrification is the secondary relaxation due to the local molecular motions in the glassy states \cite{CrystallizationModel1,CrystallizationModel2}. Motion of the entire molecule is requires for the crystal nucleation. The secondary relaxation of intermolecular origin called Johari Goldstein (JG) $\beta$-relaxation, which is a precursor of the structural $\alpha$-relaxation, is credited to the initial process of crystallization\cite{Ngai_Book01,NSKK_Nilutamide}.
\par
Clotrimazole is one of the most relevant active pharmaceutical for the antifungal medication. Clotrimazole is listed in world health organization model list of essential medicines\cite{WHO}. Clotrimazole is poorly water soluble with lower bioavailability\cite{Clotri_Bioavailability}. The clotrimazole drug is classified as slowly crystallizing compound\cite{TaylorClassifica_01,TaylorClassifica_02}. Structural characterisation of amorphous clotrimazole with NMR, X-ray and neutron diffraction has been studied by Benmore \textit{et al.}\cite{Structural_Clotri}. Viscosity influence of glass forming ability of clortimazole has been investigated by Baird\textit{et al.}\cite{Viscosity_Clotri}. However, no investigation is reported in literature for the molecular motion of amorphous clotrimazole. In the present study and dielectric relaxation of glassy and supercooled states of clorimazole is studied with broadband dielectric spectroscopy.

\section{Materials and Methods}
\subsection{Materials}
\begin{figure}[h!]
\includegraphics[width=40mm]{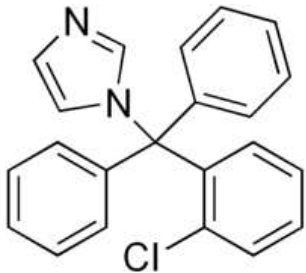}
\caption{Molecular structure of clotrimazole}
\label{fig:CLOTRI_struct}
\end{figure}

Clotrimazole (1-(2-Chlorophenyl)(diphenyl)methyl]-1H-imidazole) with chemical formula C$_{22}$H$_{17}$ClN$_{2}$ (Sigma Aldrich) in crystalline powder was used without further purification. The chemical structure of clotrimazole is shown in Fig.\ref{fig:CLOTRI_struct}.

\subsection{Methods}

\subsubsection{Raman spectroscopic measurements}
Raman spectroscopic measurement of crystalline and amorphous clotrimazole has been carried out with 785 nm radiation. Raman shift of the range of 100 to 1800 cm$^{-1}$ has been obtained. Raman spectra of crystalline (powder) sample are obtained at room temperature. Crystalline powder of the sample is heated above melting temperature (428 K) and kept 10min. to ensure complete melting. Then the melt is quenched with cooling rate of 15 K/min to a temperature of 173 K. Raman spectroscopic measurements of glass and supercooled states clotrimazole of various temperatures is obtained by heating the glass samples from 173K to 348K.  
\subsubsection{Broadband dielectric spectroscopy measurements}
The powder sample of clotrimazole is carefully filled in a stainless steel sample cell of 30mm diameter. Teflon of thickness 0.2mm was used as the spacer for the electrodes. The sample cell is heated a few degree above the melting temperature (428K) and kept 15 minutes to ensure the complete melting of the sample. Then the sample cell is pressed well and avoided forming bubbles. Then it is quenched with a rate of 15K/min to a temperature of 173K. Alpha analyser of Novocontrol broadband dielectric spectrometer was used to obtain the complex dielectric data for the frequency range of $10^{-2}$Hz to $10^{7}$Hz for various temperatures. Quatro cryosystem temperature controller of nitrogen gas cryostat is set to a temperature stability of 0.1K and used for the measurements. The glass and super cooled states of the clotimazole is obtained by heating the quenched sample from the temperature 173K to 368K in proper steps.

\section{Results and discussion}
\begin{figure}[h!]
\includegraphics[width=80mm]{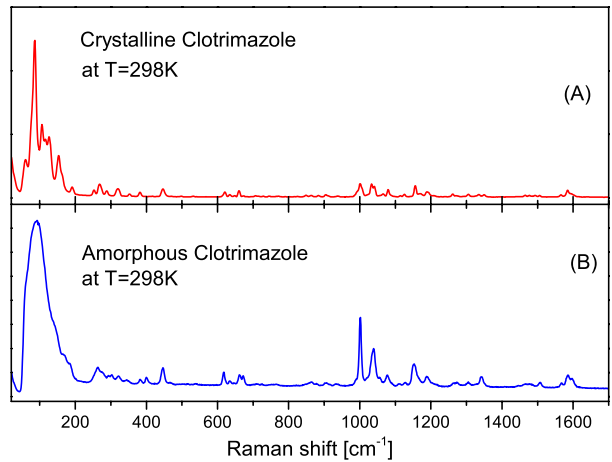}
\caption{Raman spectra of clotrimazole in (A) crystalline and (B) amorphous form.}
\label{fig:CLOTRI_Raman}
\end{figure}

The crystalline powder form of clotrimazole show all characteristic peaks of clotrimazole and it is shown in Fig.\ref{fig:CLOTRI_Raman}B. Clotrimazole show characteristic peak at 1001cm$^{-1}$, 1026cm$^{-1}$, 1156cm$^{-1}$, 1585cm$^{-1}$ as in literature\cite{Nilu_CrystEnggComm2014}. Amorphous clotrimazole show broader peaks as shown in Fig.\ref{fig:CLOTRI_Raman}A. The vibration bands related to the benzene rings 1001cm$^{-1}$, 1026cm$^{-1}$ are much more stronger compared to other peaks. Glass and supercooled states of the clotrimazole do not show typical change of position of bands. The main difference of Raman spectra of amorphous clotrimazole from the crystalline counterpart is the broadening of the band and softening and hardening of the modes. There is a possibility of formation of different conformers due to the change of orientation of planes of the benzene ring of the clotrimazole in the amorphous state.

\par
\begin{figure*}[th!]
\includegraphics[width=160mm]{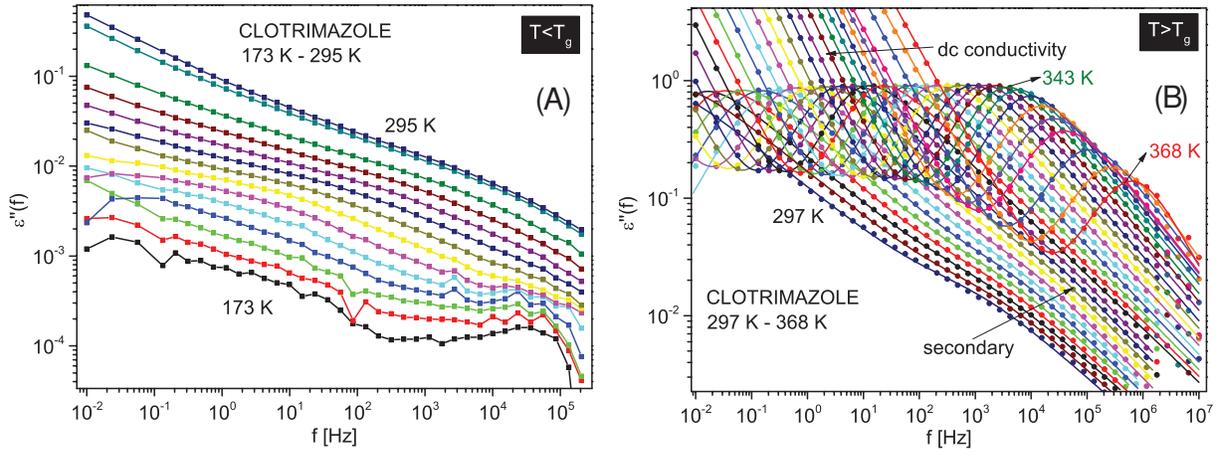}
\caption{The dielectric loss obtained for the glassy and supercooled states of clotrimazole. (A) $\varepsilon''(f)$ at temperatures below T$_{g}$ do not show well resolved secondary relaxation process (line is used for the guidance for eye). (B) $\varepsilon''(f)$ at temperatures  above T$_{g}$ show the structural $\alpha$-relaxation process and secondary relaxation process at high frequencies and dc conductivity due to the translational motion of ions. The complex dielectric data are fitted with Eq.(\ref{eqn:CLOTRI_eps_fit}) and it is shown as line.}
\label{fig:CLOTRI_eps_im}
\end{figure*}
\begin{figure}[th!]
\includegraphics[width=80mm]{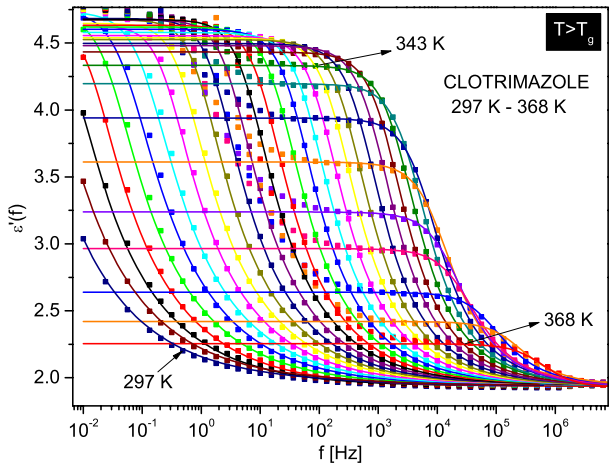}
\caption{$\varepsilon'(f)$ data of clotrimazole at temperature above T$_{g}$. Samples shows a rapid decrease of dielectric strength at temperature above 343K as an indication of crystallization.}
\label{fig:CLOTRI_eps_re}
\end{figure}

Complex dielectric data $\varepsilon^{*}(f)$=$\varepsilon'(f)$-$i \varepsilon''(f)$, of glassy and supercooled clorimazole is obtained using broadband dielectric spectroscopy. The dielectric loss data $\varepsilon''(f)$ of amorphous clotrimazole, below and above glass transition temperature (T$_{g}$) is shown in Fig.\ref{fig:CLOTRI_eps_im}A and Fig.\ref{fig:CLOTRI_eps_im}B respectively.
The $\varepsilon''(f)$ below T$_{g}$ do not show well separated secondary relaxations. The $\varepsilon''(f)$ spectra above T$_{g}$ show structural $\alpha$-relaxation and shoulder like secondary relaxation at high frequencies. Effects of dc-conductivity are seen at low frequencies of high temperature data. Two dielectric process with dc conductivity effects are generally treated as following  complex dielectric function
\begin{eqnarray}
\varepsilon^{*}(\omega)&&=-i\frac{\sigma_{dc}}{\varepsilon_{0}\omega}+\frac{\Delta\varepsilon_{\alpha}}{(1+(i\omega \tau_{\alpha_{_{HN}}})^{\alpha_{_{HN}}})^{\beta_{_{HN}}}}    \nonumber  \\
&&+\frac{\Delta\varepsilon_{_{\beta}}}{(1+(i\omega \tau_{_{\beta_{_{CC}}}})^{(1-\alpha_{_{CC}})})}+\varepsilon_{\infty}
\label{eqn:CLOTRI_eps_fit}
\end{eqnarray}
where $\sigma_{dc}$ is the dc conductivity due to the translational motion of ions. $\sigma_{dc}$ contributes $\sigma_{dc}/(\varepsilon_{0}\omega)$ to the dielectric loss. Dielectric constant has high frequency limit $\varepsilon_{\infty}$. Asymmetric function Havriliak-Negami\cite{HNoriginal1967} (HN) can be used to model the primary $\alpha$-process which is an effect of cooperative motion of clotrimazole molecule. This gives the information of the about the organization of structure of liquid leading to viscous flow. HN function characterises the $\alpha$-relaxation process with dielectric strength $\Delta\varepsilon_{\alpha}$, relaxation time $\tau_{\alpha_{_{HN}}}$ and shape parameters $\alpha_{_{HN}}$ and $\beta_{_{HN}}$. Symmetric function Cole-Cole (CC) is generally used to model the secondary process, with dielectric strength $\Delta\varepsilon_{_{\beta}}$, relaxation time $\tau_{_{\beta_{_{CC}}}}$ and shape parameter $\alpha_{cc}$. 
The complex dielectric data of clotrimazole is fitted using the Eq.(\ref{eqn:CLOTRI_eps_fit}). The dielectric loss data with fit curve is shown in Fig.\ref{fig:CLOTRI_eps_im}B and corresponding real part of complex dielectric data is shown in Fig.\ref{fig:CLOTRI_eps_re}.
The relaxation peak moves towards higher frequencies with increase of temperature as relaxation process become faster. Even though clotrimazole is classified as slowly crystallizing compound which shows no crystallization during the DSC measurements, amorphous clotrimazole crystallizes above the temperature of 343K during dielectric measurements. The molecule may acquire sufficient energy at this temperature to crystallize. Rapid decrease of dielectric strength of structural relaxation process is occurred above 343K as an indication of crystallization and it is shown in Fig.\ref{fig:CLOTRI_eps_re}. The density of the mobile dipoles decreases as the increase of crystallization and hence the strength of the structural relaxation. Isotherm dielectric measurement with time may give some insight to the crystal growth of the amorphous pharmaceuticals\cite{DrugDelRev2015,NSKK_Nilutamide}. 
\par
\begin{figure}[h]
\centering
\includegraphics[width=70mm]{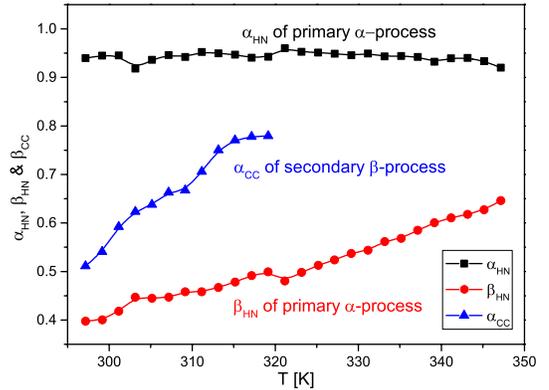}
\caption{Temperature dependence of the shape parameters $\alpha_{_{HN}}$ and $\beta_{_{HN}}$ of the primary process and $\alpha_{cc}$ of the secondary relaxation. The $\alpha_{_{HN}}$ is weakly temperature dependent and has a value nearly one. The $\beta_{_{HN}}$ and $\alpha_{cc}$ increases with increase of temperature. It indicates that both primary and secondary relaxation process of clotrimazole tends towards Debye with increase of temperature.}
\label{fig:CLOTRI_fitpara}
\end{figure}
\begin{figure}[h]
\centering
\includegraphics[width=60mm]{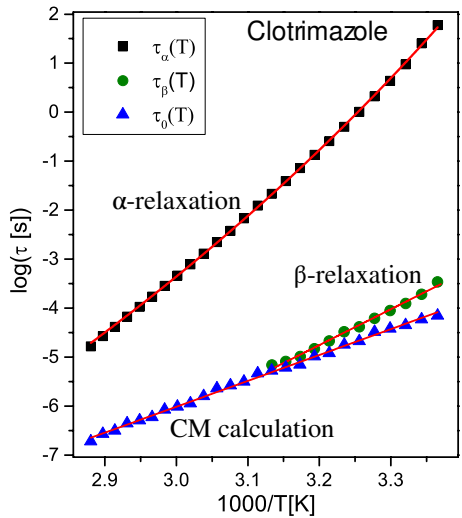}
\caption{Thermal activation of $\alpha$-relaxation and $\beta$-relaxation processes of clotrimazole. Primitive relaxation $\tau_{_{0}}(T)$ calculated by coupling model is shown.}
\label{fig:CLOTRI_vft}
\end{figure}
\begin{figure}[h!]
\centering
\includegraphics[width=80mm]{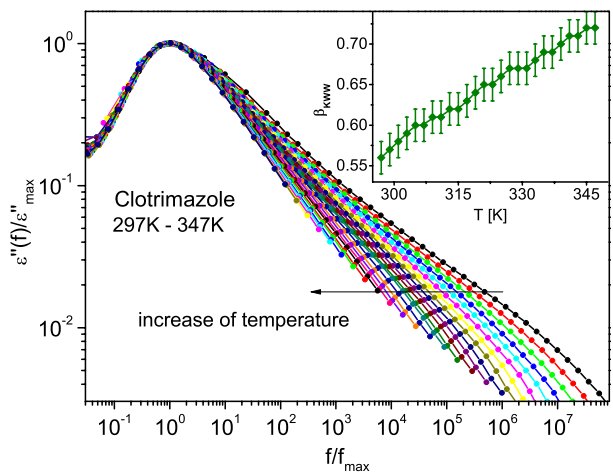}
\caption{The scaled dielectric loss of clotrimazole at different temperatures do not overlap each other to produce the master curve. It indicates temperature dependence of $\beta_{KWW}$ and it is shown in the inset.}
\label{fig:CLOTRI_scaling}
\end{figure}

The shape parameters $\alpha_{_{HN}}$ and $\beta_{_{HN}}$ of the structural relaxation process and the $\alpha_{cc}$ of the secondary relaxation process show interesting temperature dependants as in Fig.\ref{fig:CLOTRI_fitpara}.
The exponent $\alpha_{_{HN}}$ behaves almost temperature independent with value nearly one, which indicates Cole-Davison type behaviour of $\alpha$-relaxation process. While the values of $\beta_{_{HN}}$ of the $\alpha$-relaxation process, increase with increase of temperate and its value ranges from 0.40 to 0.65. The shape parameter $\alpha_{cc}$ of the secondary relaxation show more temperature dependence and values ranges from 0.51 to 0.78. This indicates both the primary and secondary relaxation tending towards the Debye dielectric relaxation with increase of temperature.
The structural relaxation time $\tau_{\alpha}$ is obtained from the peak maximum frequency. It can be calculated from the parameters of HN function \cite{RichertHNPeak1994}.
\begin{equation}
\tau_{\alpha}=\tau_{\alpha_{_{HN}}}sin\left[\frac{\alpha_{_{HN}}\pi}{2+2\beta_{_{HN}}}\right]^{-1/\alpha_{_{HN}}}sin\left[\frac{\alpha_{_{HN}}\beta_{_{HN}}\pi}{2+2\beta_{_{HN}}}\right]^{1/\alpha_{_{HN}}}
\label{eqn:tau_max}
\end{equation}
The relaxation time and viscosity of the supercooled liquids generally show non-Arrhenius behaviour. Thermal activation of $\alpha$-process of clotrimazole is shown in Fig.\ref{fig:CLOTRI_vft}.
The relaxation time $\tau_{\alpha}$ follows empirical Vogel-Fulcher-Tammann (VFT) equation\cite{VFT1-1921,VFT2-1923,VFT3-1926}.
\begin{equation}
\tau_{\alpha}=\tau_{\infty} exp[D T_{0}/(T-T_{0})]
\label{eqn:CLOTRI_VFT}
\end{equation}
The strength parameter D=31.1$\pm$0.9, log($\tau_{\infty}$)=-19.5$\pm$0.5 and T$_{0}$=181.9$\pm$2.8K is obtained. It is widely considered that at the glass transition temperature T$_{g}$, the structural relaxation time, $\tau_{\alpha}$=100s and for clotrimazole T$_{g}$=296$\pm$11 is obtained from VFT parameters. $T_{g}$ obtained from dielectric data is a few degree different from T$_{g}$=303K obtained from the DSC measurement by Baird\textit{et al.}\cite{Viscosity_Clotri}, but it is within the range of experimental error. Glasses which has Arrhenius beahviour of thermal activation for the structural relaxation are referred as ``strong'' and with non-Arrhenius beahviour are termed ``fragile''. The amount of deviation of relaxation time from Arrhenius is generally quantifies with the fragility index obtained from the Angell plot\cite{Angell_fragility}. The fragility index m, can also be obtained from the VFT parameters as
\begin{equation}
m=\left.\frac{dlog_{_{10}}\tau_{\alpha}}{d(T_{g}/T)}\right|_{T=T_{g}}=D\frac{T_{0}}{T_{g}}\left(1-\frac{T_{0}}{T_{g}}\right)^{-2}log_{_{10}}e
\label{eqn:VFT}
\end{equation}
Cooling below glass transition temperate arrests the liquid structure of amorphous material and the fragility index describe how rapidly the arrested liquid structure disrupted on heating. Higher the fragility higher is the glass forming ability and physical stability of amorphous system. The fragility is a key factor to find the appropriate storage condition of particular amorphous drug. The fragility index of clotrimazole is m=50$\pm$12, and it can be compared with many other glass formers, 2,3-dimethylpentane (m=50), 3-bromopentane (m=53), butyronitrile (m=47). Clotrimazole can be classified as the intermediate glass-former drug. Almost all glass-forming materials have non-exponential molecular relaxation process and it is modelled with Kohlrausch-Williams-Watts (KWW) stretched exponential function.
\begin{equation}
\phi(t)=exp\left[-\left(\frac{t}{\tau_{_{KWW}}}\right)^{\beta_{_{KWW}}}\right]
\label{eqn:KWW}
\end{equation}
The time domain function KWW do not have a analytical function in frequency domain for all values of exponent $\beta_{KWW}$. The frequency domain data of the KWW function can be obtained by numerical calculation of one sided Fourier transform. The scaled dielectric loss of clotrimazole do not overlap each other to produce master curve as shown in Fig.\ref{fig:CLOTRI_scaling}.
It indicates the stretched exponential exponent $\beta_{KWW}$ is temperature dependent. The exponent $\beta_{KWW}$ increases with increase of temperature. Its value ranges from 0.56 to 0.72 and it is shown in inset of Fig.\ref{fig:CLOTRI_scaling}. This show that the relaxation process tending towards Debye process with increase of temperature as indicated by the shape parameters obtained from Eq.(\ref{eqn:CLOTRI_eps_fit}). The KWW exponent $\beta_{KWW}$=0.56 near $T_{g}$ and it is comparable with many other pharmaceutical materials Ibuprofen (0.55), Indometacin (0.59) and other glass forming materials 4-methylheptane (0.57), tetraphenyl tetramethyl trisiloxane (0.58), Tributylacetyl citrate (0.57), Trimethyl pentaphenyl trisiloxane (0.57) and Triphenyl phosphate (0.56). 
\par
Shamblin \textit{et al.} observed that materials of lower the value of $\beta_{KWW}$ have higher crystal nucleation rate\cite{Shamblin_KWW_Nucleation}. The distribution of structural relaxation time reduces the physical and chemical stability and consequently the shelf life of the pharmaceutical. Faster modes of molecular motions within the distribution of relaxation time can be responsible for the nucleation in the glassy state, therefore the glass formers with smaller $\beta_{KWW}$ would be more susceptible to nucleate\cite{Johari_KWW_Nucleation}. But this correlation is not shown by many of the glass forming pharmaceuticals\cite{DrugDelRev2015}. Bohmer et.al. suggested an empirical relation m=250$\pm$30-320$\beta_{KWW}$ and it follows most of the glass pharmaceuticals\cite{Bohmer_KWW-m}. Clotrimazole also follows this empirical relation. The stretching parameter $\beta_{KWW}$ has been often considered as a measure of degree of cooperativity and length scale or dynamic heterogeneity of molecular mobility reflected in the structural relaxation\cite{DrugDelRev2015}. 
\par 
Paluch and co-workers observed that width of $\alpha$-loss peak near T$_{g}$ of the van der Waals glass formers strongly anticorrelates with the dielectric strength. For large vales of $\Delta\varepsilon_{\alpha}$ larger the $\beta_{KWW}$\cite{Ngai_eps_KWW_PRL_2016}. Clotrimazole has $\Delta\varepsilon_{\alpha}$=2.7$\pm$0.2 and $\beta_{KWW}$=0.56$\pm$0.02 near $T_{g}$, which also follows this anticorrelation as intermediate to Indometacin($\Delta\varepsilon_{\alpha}$=3.8, $\beta_{KWW}$=0.59) and Itraconazol($\Delta\varepsilon_{\alpha}$=2.4, $\beta_{KWW}$=0.50) behaviour. The contribution of dipole-dipole interaction potential, V$_{dd}\propto$r$^{-6}$ (r is the distance between the molecules), to the attractive part of the intermolecular potential which makes the resultant potential more harmonic and the dispersion of $\alpha$-relaxation narrower. 
\par
\begin{figure}[h!]
\centering
\includegraphics[width=90mm]{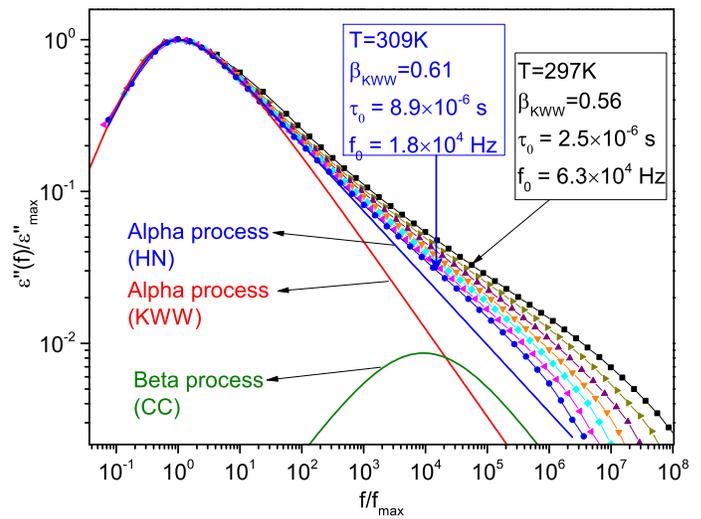}
\caption{The scaled $\varepsilon''(f)$ at T=309K with KWW fit (red curve),  HN function fit for $\alpha$-process (blue curve) and CC function fit for $\beta$-relaxation process (green curve). The primitive relaxation time $\tau_{_{0}}$ obtained from the coupling model corresponds a frequency f$_{0}$ and the secondary process are close to each other showing the secondary relaxation process in clotrimazole is Johar-Goldstein $\beta$-relaxation process. Primitive relaxation frequency f$_{0}$ for T=297K is also shown in the figure.}
\label{fig:CLOTRI_KWWCoupling}
\end{figure}

It is really challenging to find the origin of secondary relaxation process in glass forming materials. Secondary relaxation process may originates either from  inter-molecular or intramoleacular process. Intermolecular process with the entire motion of the molecule may produce a secondary relaxation process called Johari-Goldstein (JG) $\beta$-process, which is precursors of the primary structural relaxation process\cite{JG,Ngai_Beta_classification}. The JG process is considered to be a universal process for all glass formers and fast secondary relaxation process of intramolecular origin are generally noted as $\gamma$-process. Coupling model gives the correlation of $\alpha$ and $\beta$ processes\cite{Ngai_Coupling1,Ngai_Coupling2}. The primitive relaxation of $\tau_{_{0}}$ can be obtained from the following empirical relation.
\begin{equation}
\tau_{_{0}}=(t_{c})^{n}(\tau_{\alpha})^{(1-n)}
\label{eqn:Ngaicoupling}
\end{equation}
where n=1-$\beta_{KWW}$ is the coupling parameter, which is obtained from the structural $\alpha$ relaxation process. The time characterizing the crossover from independent to cooperative fluctuation t$_{c}$ has vale 2$\times$10$^{-12}$s for small glassformers. The scaled dielectric loss data with the calculated primitive relaxation at temperatures 297K and 309K are shown in the Fig.\ref{fig:CLOTRI_KWWCoupling}.
The scaled dielectric loss data at 309K is fitted with the Eq.(\ref{eqn:CLOTRI_eps_fit}). The individual contributions of the $\alpha$-process (HN) and $\beta$-process (CC) relaxation process are shown. The $\beta_{KWW}$=0.61 is obtained for T=309K and primitive relaxation time $\tau_{_{0}}$=8.89$\times10^{-6}$s obtained. Since Cole-Cole function is symmetric $\tau_{_{\beta_{_{CC}}}}$ and relaxation time from peak maximum frequency are same and it is referred as $\tau_{_{\beta}}$. The secondary relaxation time $\tau_{_{\beta}}$ is well within the calculated primitive relaxation and hence the secondary relaxation of clotrimazole may be considered as the Johar-Goldstein $\beta$-relaxation. The primitive relaxation calculated for the temperature T=297K with $\beta_{KWW}$=0.56 also shown in the Fig.\ref{fig:CLOTRI_KWWCoupling}. The calculated primitive relaxation $\tau_{_{0}}$ and secondary relaxation $\tau_{cc}$ from Eq.(\ref{eqn:CLOTRI_eps_fit}), is shown in the Fig.\ref{fig:CLOTRI_vft}. The $\beta$-relaxation process is thermally activated and follows Arrhenius equation.
\begin{equation}
\tau_{_{\beta}}=\tau_{\infty} exp[E_{_{\beta}}/k_{_{B}}T]
\label{eqn:Arrhenius_time}
\end{equation}
The activation energy obtained from $\tau_{cc}$ is about 139.9 kJ/mol and that from $\tau_{_{0}}$ is about 101.2 kJ/mol. The dielectric strength of $\beta$-process, $\Delta\varepsilon_{\beta}$ decreases with decrease of temperature. This can be explained in terms of the decrease in number of islands of mobility with decrease of temperature. Near the temperature T$_{g}$, $\beta$-process is more close to $\alpha$-process, with reduced dielectric strength $\Delta\varepsilon_{\beta}$ as compared to that at higher temperatures.

\section{Conclusion}
One of the relevant pharmaceutical drug clotrimazole melt quenched and the molecular dynamics of glassy and supercooled states are studied using dielectric spectroscopy. Complex dielectric data of amorphous clotrimazole has been obtained from dielectric spectroscopy for a wide range of temperature and frequency. The dielectric data show structural relaxation, $\alpha$-process which follows VFT equation. Glass transition temperature T$_{g}$ and fragility m, is obtained from the VFT parameters. Clotrimazole is considered to be an intermediate fragile glass former. The dielectric data do not show a well separated secondary relaxation. The $\alpha$-process at high frequencies are influenced by the secondary relaxation process and shoulder like shape is formed. This secondary relaxation process coincides with the coupling model predicted primitive relaxation. Hence the secondary relaxation process in the clotrimazole is JG $\beta$-relaxation. The KWW stretched exponent $\beta_{KWW}$=0.56, near glass transition temperatures. The value of $\beta_{KWW}$ increases with increase of temperature and hence the scaled dielectric loss do not form a single master curve. 
\begin{acknowledgments}
Authors thank Central Instrumentation Facility, Pondicherry University for the dielectric spectroscopy measurements. Authors acknowledge CSIR project F.03(1279)/13/EMR-II for financial support and senior research fellowship for the author NSKK.
\end{acknowledgments}

\section*{References}
\bibliography{clotri}
\end{document}